\documentclass{Rinton-P9x6}

\begin{document}

\title{Efficient classical simulation of measurements in optical
  quantum information}

\author{Stephen D. Bartlett and Barry C. Sanders}

\address{Department of Physics and Centre for Advanced Computing --
  Algorithms and Cryptography, Macquarie University, Sydney, NSW 2109,
  Australia 
  \\E-mail: stephen.bartlett@mq.edu.au
}


\maketitle

\abstracts{We present conditions for the efficient simulation of a
  broad class of optical quantum circuits on a classical machine: this
  class includes unitary transformations, amplification, noise, and
  measurements.  Various proposed schemes for universal quantum
  computation using optics are assessed against these conditions, and
  we consider the minimum resource requirements needed in any optical
  scheme to generate optical nonlinear processes and perform universal
  quantum computation.}


Information processing using the rules of quantum mechanics may allow
tasks that cannot be performed using classical laws\cite{Nie00}.  Of
the many possible realizations of quantum information processes,
optical realizations have the advantange of negligible decoherence.
Both qubit\cite{Chu95,KLM01,Got01b} and
continuous-variable\cite{Llo99} schemes offer significant potential
for optical quantum information processing, especially if efficient
processes can be performed that are not efficient on any classical
device.

Advanced techniques in linear optics and squeezing are known to be
insufficient to perform universal quantum
computation\cite{Nie00,Llo99,Bar02a}; in particular, optical nonlinear
processes (such as a Kerr nonlinearity\cite{Wal94}) have been
identified as a necessary requirement.  (Schemes that employ only
linear optics\cite{Cer98} are not scalable, in that they require
resources that grow exponentially in the number of qubits.)  However,
Kerr nonlinearities suffer either from weak strengths or high losses,
and the lack of appropriate nonlinear materials greatly restricts the
type of processes that can be performed in practice.

Recently, nonunitary processes such as measurement have been
identified as a means to implement nonlinear operations.  Proposals
for optical quantum computation by Knill, Laflamme and
Milburn\cite{KLM01} (KLM) and Gottesman, Kitaev and
Preskill\cite{Got01b} (GKP) employ photon counting to induce nonlinear
transformations in optical systems.  Photon counting is an important
example of a process that can be used to achieve nonlinear
transformations via feedforward of measurement results.  Such a
nonunitary transformation appears to enable impressive capabilities
equivalent to nonlinear transformations.

It is imperative to determine what type of processes (unitary
transformations, projective measurements, interaction with a
reservoir, etc.)\ can be used to implement nonlinear transformations
and thus perform universal quantum computation.  One approach is to
identify classes of processes that can be efficiently simulated on a
classical computer.  Under the assumption that universal quantum
computation is \emph{not} efficiently simulatable classically, such
processes are also insufficient to implement optical nonlinear
transformations.  The Gottesman-Knill (GK) theorem\cite{Nie00} for
qubits, the continuous-variable classical simulatability theorem of
Bartlett \emph{et al}.\cite{Bar02a} (BSBN), and the general optical
classical simulatability theorem of Bartlett and Sanders\cite{Bar02b}
(BS) allow us to investigate the classical complexity of a quantum
process using particular classes of initial states, unitary operations
and measurements.

Here, we consider the implications of the BS optical classical
simulatability theorem on various proposals for quantum computation
using optics.  This theorem employs the powerful formalism of Gaussian
completely positive (CP) maps\cite{Lin00} to describe efficiently
simulatable operations on Gaussian states.

\medskip
\noindent \textbf{Theorem for efficient classical simulatability
  (BS):} \textit{Any quantum information process that initiates in a
  Gaussian state and that performs only Gaussian CP maps can be
  \emph{efficiently} simulated using a classical computer.}  These
maps include (i) the unitary transformations corresponding to linear
optics and squeezing, (ii) linear amplification (including
phase-insensitive and phase-sensitive amplification and optimal
cloning), linear loss mechanisms or additive noise, (iii) measurements
that are Gaussian CP maps including, but not limited to, projective
measurements in the position/momentum eigenstate basis or
coherent/squeezed state basis, with finite losses, and (iv) any of the
above Gaussian CP maps conditioned on classical numbers or the
outcomes of prior Gaussian CP measurements (classical feedforward).
\medskip

Our theorem for efficient classical simulation provides a powerful
tool in assessing whether a given optical process can enhance linear
optics to perform nonlinear transformations or allow quantum processes
that are exponentially faster than classical ones.  Algorithms or
circuits employing Gaussian CP maps can be efficiently simulated on a
classical computer, and thus do not provide any sort of quantum
exponential speedup.


We now consider some of the key new results of this theorem in terms
of known processes.  

\medskip
\noindent \textbf{Corollary 1:}  Linear optics or squeezing
transformations conditioned on the measurement outcome of homodyne
detection with finite losses using Gaussian states cannot induce a
nonlinearity.  
\medskip

Thus, initiating with Gaussian states, it is not possible to use
homodyne measurements and feedforward of measurement results to induce
a (possibly nondeterministic) optical nonlinearity in the way that
photon counting allows in the KLM scheme.  In terms of optical
implementations of quantum computing, this theorem reveals why all
previous schemes either propose some form of optical
nonlinearity\cite{Chu95,Llo99}, use other forms of measurement such as
photon counting\cite{KLM01,Got01b} or are not efficiently
scalable\cite{Cer98}.

This theorem also places severe constraints on the use of
photodetection to perform nonlinear transformations in realizations of
optical quantum computing.  For a threshold
photodetector\cite{KLM01,Kok01,Bar02c} with perfect efficiency, the
POVM is given by two elements, corresponding to ``absorption'' and
``no-absorption'' of light.  Photon counters are effectively
constructed as arrays of such detectors\cite{Bar02c}.  The vacuum
projection describes the non-absorption measurement, and the
corresponding map describing this measurement result is Gaussian CP.
However, the absorption outcome is not.

\medskip
\noindent \textbf{Corollary 2:}  Gaussian CP maps conditioned on the
no-absorption outcome of a photodetection measurement are also
Gaussian CP, whereas transformations conditioned on the absorption
outcome are not.  \medskip

Note that the same result holds for finite-efficiency photodetectors:
such detectors can be modelled as unit efficiency photodetectors
with a linear loss mechanism describable using Gaussian CP maps.
Thus, the absorption outcome of photodetection and the feedforward of
this measurement result is a key resource for optical quantum
information processing.  This corollary also proves that any nonlinear
gate employing linear optics and photon counting \emph{must} be
nondeterministic; a photon counting measurement of a Gaussian state
could possibly result in an outcome of zero photons, and such a result
corresponds to an efficiently classically simulatable process.  (Note
that nonlinear optics, in contrast, can be deterministic.)

\begin{table}[t]
\caption{Efficient classical simulatability for schemes employing
  various initial states, unitary gates, and measurements.\label{tab:Sim}}
\begin{center}
\footnotesize
\begin{tabular}{|l|l|l|l|}
\hline
{\bf Initial States} &\raisebox{0pt}[13pt][7pt]{\bf Unitary Gates} &
\raisebox{0pt}[13pt][7pt]{\bf Measurements} & 
\begin{minipage}{0.7in}{\bf Efficiently}\\{\bf simulatable?}
\end{minipage}\\
\hline
{Vacua} &\raisebox{0pt}[13pt][7pt]{Linear optics, squeezing} &
\begin{minipage}{1in}
{Gaussian CP}\\ {(i.e., homodyne)}
\end{minipage} &{Yes\cite{Bar02a,Bar02b}}\\
\hline
{Vacua} &\begin{minipage}{1.25in}{Linear optics, squeezing,}\\ {Kerr
    nonlinearity} \end{minipage} &
\raisebox{0pt}[13pt][7pt]{Homodyne} &
\begin{minipage}{0.7in}{No (Lloyd \&}\\  {Braunstein\cite{Llo99})}
\end{minipage}\\
\hline
{Single photons} &\raisebox{0pt}[13pt][7pt]{Linear optics only} &
\raisebox{0pt}[13pt][7pt]{Photon counting} &
\raisebox{0pt}[13pt][7pt]{No (KLM\cite{KLM01})}\\
\hline
{Vacua} &\raisebox{0pt}[13pt][7pt]{Linear optics, squeezing} &
\begin{minipage}{1in}{Photon counting}\\{\& homodyne} 
\end{minipage}&
\raisebox{0pt}[13pt][7pt]{No (GKP\cite{Got01b})}\\
\hline
{Single photons} &\raisebox{0pt}[13pt][7pt]{Linear optics, squeezing} &
\raisebox{0pt}[13pt][7pt]{Homodyne} &{???}\\
\hline
\end{tabular}
\end{center}
\end{table}

Our classical simulatability theorem may be useful in assessing the
``minimum'' requirements for universal quantum computation with
optics.  Table 1 presents various classes of initial states, unitary
gates, and measurements (that can be used for classical feedforward)
and their classical simulatability according to our theorem.
Employing only Gaussian states and Gaussian CP maps results in an
efficiently simulatable circuit; one can now consider supplementing
this set with various ``resources'' that may allow for universal
quantum computation.  As shown by Lloyd and Braunstein\cite{Llo99},
the addition of a Kerr nonlinearity or any higher-order transformation
on a single mode results in universal quantum computation.  The
schemes of KLM and GKP reveal that photon counting is also a resource
that allows for universal quantum computation.  The KLM scheme also
requires single photon Fock states ``on demand'' as ancilla inputs to
their nondeterministic nonlinear gates; such states lie outside the
domain of our theorem (they are not Gaussian) and may serve as a
resource for performing nonlinear operations.  

It is interesting to consider, then, if single photons on demand are
by themselves sufficient to bestow Gaussian CP maps with the power to
perform nonlinear operations and thus universal quantum computation.
Considering the recent progress in creating single photon turnstile
devices\cite{San01} (with low probablility of producing zero or two
photons by accident), a scheme that requires single photons but
otherwise employs only linear optics, squeezing, and high-efficiency
homodyne detection would obviate the need for ultra-high efficiency
photon counters\cite{Bar02d}.

\section*{Acknowledgments}

SDB acknowledges helpful discussions with M.\ A.\ Nielsen.  This project
has been supported by an ARC Large Grant and a Macquarie University
New Staff Grant.

\end{document}